\RequirePackage{fix-cm}
\documentclass[smallcondensed]{svjour3}     % onecolumn (ditto)
\smartqed  % flush right qed marks, e.g. at end of proof
\usepackage{graphicx}
\usepackage{url}
\usepackage{hyperref}
\usepackage[numbers,sort&compress]{natbib}

\usepackage[labeled,resetlabels]{multibib}
\newcites{W}{URL references and permalinks}
\usepackage{xparse}
\let\oriCiteW\citeW
\RenewDocumentCommand{\citeW}{O{} O{} m}{%
  \renewcommand{\citenumfont}[1]{W##1}%  
  \oriCiteW[#1][#2]{#3}%
  \renewcommand{\citenumfont}[1]{##1}%
}

\usepackage{xspace}
\usepackage{svg}
\usepackage{doi}
\usepackage[skipbelow=10pt,skipabove=10pt,innerbottommargin=10pt,innertopmargin=10pt]{mdframed}
\usepackage{multicol}
\usepackage[misc,geometry]{ifsym}
\usepackage[inline]{enumitem}

\newcounter{bluebox}
\newenvironment{bluebox}[1]{%
\begin{mdframed}[backgroundcolor=blue!3,hidealllines=true]%
\refstepcounter{bluebox}
\noindent\colorbox{blue!6}{\makebox[0.98\textwidth]{\sffamily Box~\thebluebox: #1}} \\
\small\sffamily%
}{%
\end{mdframed}
}%

\newcommand{\cpp}{C\nolinebreak\hspace{-.05em}\raisebox{.4ex}{\tiny\bf +}\nolinebreak\hspace{-.10em}\raisebox{.4ex}{\tiny\bf +}\xspace}
\newcommand\dumux{DuMu\textsuperscript{x}\xspace}

\newcommand{\eg}{e.g.\xspace}

\usepackage[draft,commentmarkup=uwave]{changes}% defaultcolor=blue
\definechangesauthor[name=timo, color=orange]{tk}
\definechangesauthor[name=anett, color=red]{as}
\definechangesauthor[name=bernd, color=blue]{bf}
\definechangesauthor[name=dennis, color=cyan]{dg}
\definechangesauthor[name=sibylle, color=magenta]{sh}
\definechangesauthor[name=katharina, color=green]{kg}
\definechangesauthor[name=david, color=brown]{db}
\definechangesauthor[name=sarbani, color=teal]{sr}

\begin{document}

\title{A sustainable infrastructure concept for improved accessibility, reusability, and archival of research software}

\titlerunning{Accessibility, reusability, and archival of research software}        % if too long for running head

\author{Timo Koch          \and
        Dennis Gläser       \and
        Anett Seeland       \and
        Sarbani Roy         \and
        Katharina Schulze    \and
        Kilian Weishaupt \and
        David Boehringer    \and
        Sibylle Hermann      \and
        Bernd Flemisch
}


\institute{T. Koch \at Department of Mathematics, University of Oslo, Norway        
           \and
           T. Koch \and D. Gläser \and S. Roy \and K. Weishaupt \and B. Flemisch (\Letter) \at Department of Hydromechanics and Modelling of Hydrosystems, IWS, University of Stuttgart, Germany \\\email{bernd.flemisch@iws.uni-stuttgart.de}
           \and A. Seeland \and K. Schulze \and D. Boehringer \at  Technical information and communication services, University of Stuttgart, Germany
           \and A. Seeland \and K. Schulze \and S. Hermann \at University Library, University of Stuttgart, Germany%
}

\date{Received: date / Accepted: date}

% Final fixes: adapted abstract to include FAIR
% tried to avoid multi-model data representation > multi-modal representation
% distinguished between reserach software and frameworks more explicitly
% Moved reproduciblilty and FAIR from chapter 2 up into the introduction
% Include multi-modal artifcats in the caption of figure 1

\maketitle

\begin{abstract}
Research software is an integral part of most research today and it is widely accepted that research software artifacts should be accessible and reproducible. However, the sustainable archival of research software artifacts is an ongoing effort. We identify research software artifacts as snapshots of the current state of research and an integral part of a sustainable cycle of software development, research, and publication. We develop requirements and recommendations to improve the archival, access, and reuse of research software artifacts based on installable, configurable, extensible research software, and sustainable public open-access infrastructure. The described goal is to enable the reuse and exploration of research software beyond published research results, in parallel with reproducibility efforts, and in line with the FAIR principles for data and software. Research software artifacts can be reused in varying scenarios. To this end, we design a multi-modal representation concept supporting multiple reuse scenarios. We identify types of research software artifacts that can be viewed as different modes of the same software-based research result, for example, installation-free configurable browser-based apps to containerized environments, descriptions in journal publications and software documentation, or source code with installation instructions. We discuss how the sustainability and reuse of research software are enhanced or enabled by a suitable archive infrastructure.
Finally, at the example of a pilot project at the University of Stuttgart, Germany---a collaborative effort between research software developers and infrastructure providers---we outline practical challenges and experiences.

\end{abstract}

%\added[id=bf]{Example for added}.\\
%Example for \replaced[id=bf]{replaced}{original}.\\
%\deleted[id=bf]{Example for deleted}.\\
%\comment[id=bf]{Example for comment}.

\section{Introduction}

Computation and the use of software are ubiquitous in most scientific disciplines~\citep{Hey2009} and play an essential role in most scientific research~\citep{Howison2015,Hettrick2018}. Many scientists write their own software~\citep{Brett2017RSE}. Research software is often an essential component of research and can also itself be the focus of the research. This leads to the questions, ``How is research software, in its state that produced the published results, best archived with high potential for accessibility and reusability?'', and ``How can the archival of \textit{research software artifacts}\footnote{Here and throughout the article, we refer with the term \textit{research software artifacts} broadly to all digital assets related to a research software reuse scenario. This includes, for example, source code, documentation, journal publications describing the software, installation scripts, container recipes and images, workflow descriptions, input and output data, etc.} promote sustainable research software development?''.

% paragraph about properties/challenges -> sustainability
\Citet{Mangul2019} recently summarized some major practical challenges with the reuse, installation, and archival of research software based on a study of (research) bioinformatics software~\citep{Mangul2019b}.
They investigated links to software artifacts in over $30\,000$ papers (published between 2005 to 2017) and found that about one-fourth of the mentioned URLs\footnote{URL stands for Uniform Resource Locator and is a type of Uniform Resource Identifier (URI) which is not necessarily a persistent identifier.} were not accessible. Moreover, only half of the $98$ randomly selected software packages in these papers were given the label ``easy install''. Installability was positively correlated with the number of citations the software publication received.
While being accessible (retrievable) and installable (deployable) are necessary properties of sustainably archived software artifacts, the community goes well beyond~\cite{Wilkinson2017}. Equally important aspects for the knowledge advance in the scientific community are the reproducibility (looking back) and reusability and extensibility of research software artifacts (looking forward).

% reproducible research
The idea of \emph{fully} reproducible research is not new---\citet{Claerbout1992} argued in the 1990s that software and reproduction instructions for all data and figures need to be published alongside classical scientific journal publications. Although most basic technical challenges have been resolved today, practical reproducibility remains a real concern, many even speaking of a reproducibility crisis~\citep{Fanelli2018}.
Even if a result is \textit{technically} reproducible (i.e. reproducible given unlimited resources), it might not be \textit{practically} reproducible by others. The reason may be as simple as that the peer lacks knowledge about a particular programming language used.
In the following, we address the aspect of \emph{practical} reproducibility. Particularly, we make suggestions for archiving research software in a way that increases the likelihood of practical reproducibility in the future.

% FAIR
In the ideal case\textemdash and in line with the FAIR principles~\citep{Wilkinson2016FAIR,Hasselbring2020,Barker2022}\textemdash research artifacts (data and software) can be (practically) \emph{explored} beyond the already published aspects. That means, for instance, varying parameter settings that the research software supports vary but have not been varied in the scope of the presented published dataset. Another example is the creation of alternative visualizations of output data.

% explorable research
Both \emph{explorability} and \emph{reusability} are desirable on their own, even if the reproducibility of a particular result cannot be achieved~\cite{Albers2017}. Moreover, failure of reproducibility may be restored if the software artifact can be reused and derived in order to remove its defects. (This requires open data and software licenses~\cite{Stodden2009}, see Section~\ref{sec:floss}.)

% effort as a major problem
A significant effort is often associated with the sustainable archival of research software artifacts. Limited resources (time and skills) of research software developers and the lack of incentives in the academic research environment to assure the quality of research software are often suggested as the main problems leading to a lack of software sustainability and software quality assurance~\citep{Kempf2017,Mangul2019}. Research software is often not reusable because of missing documentation~\cite{hermann2022documenting}, broken links, or missing data~\cite{Howison2015}. However, in a survey on reproducibility, Elmenreich et al.~\cite{ElmenreichMTL19} found among more than $100$ computer scientists, that researchers are both willing to spend time to reproduce their peer's research software artifacts (23 hours on average) \emph{and} willing to spend time to make their own research reproducible by others (24 hours on average).

% effort solved by automating workflows
The effort for sustainable archival of research artifacts can be significantly reduced by workflow automation. One example is the use of workflow tools\footnote{A list of workflow environments has been compiled in~\citeW{awesome-pipe}. \cite{Diercks2022Preprint} review requirements of workflow tools in the context of sustainable scientific workflows and compare various tools.}. %designed to manage reproduction steps and workflow metadata with a level of abstraction beyond what is possible with most basic scripting tools (e.g. \texttt{bash}) or more advanced build environments (e.g. \texttt{CMake}).
Virtualization or containerization makes it possible to install and develop software in reproducible environments. If installation recipes, including software dependencies, are part of the workflow record, containerization can be automated~\cite{Repo2Docker2018,containerit2019}. However, creating installation recipes to reproduce the correct runtime environment is challenging and usually a domain-specific process.

% infrastructure automation, metadata
% reuse and exploration
Research software artifacts have to be preserved to enable the reproduction of scientific results. This requires infrastructure based on free and open-source software (see Section~\ref{sec:researach-software})
%, efficient and flexible archive workflow tools, 
and suitable metadata schemes and their automated extraction from workflows in compliance with the FAIR principles~\cite{Wilkinson2016FAIR,Barker2022} (findable, accessible, interoperable, reusable). The FAIR principles state goals but stay deliberately open on how these goals are to be achieved.

\begin{figure}[!htbp]
    \centering
    \includegraphics[width=1.0\textwidth]{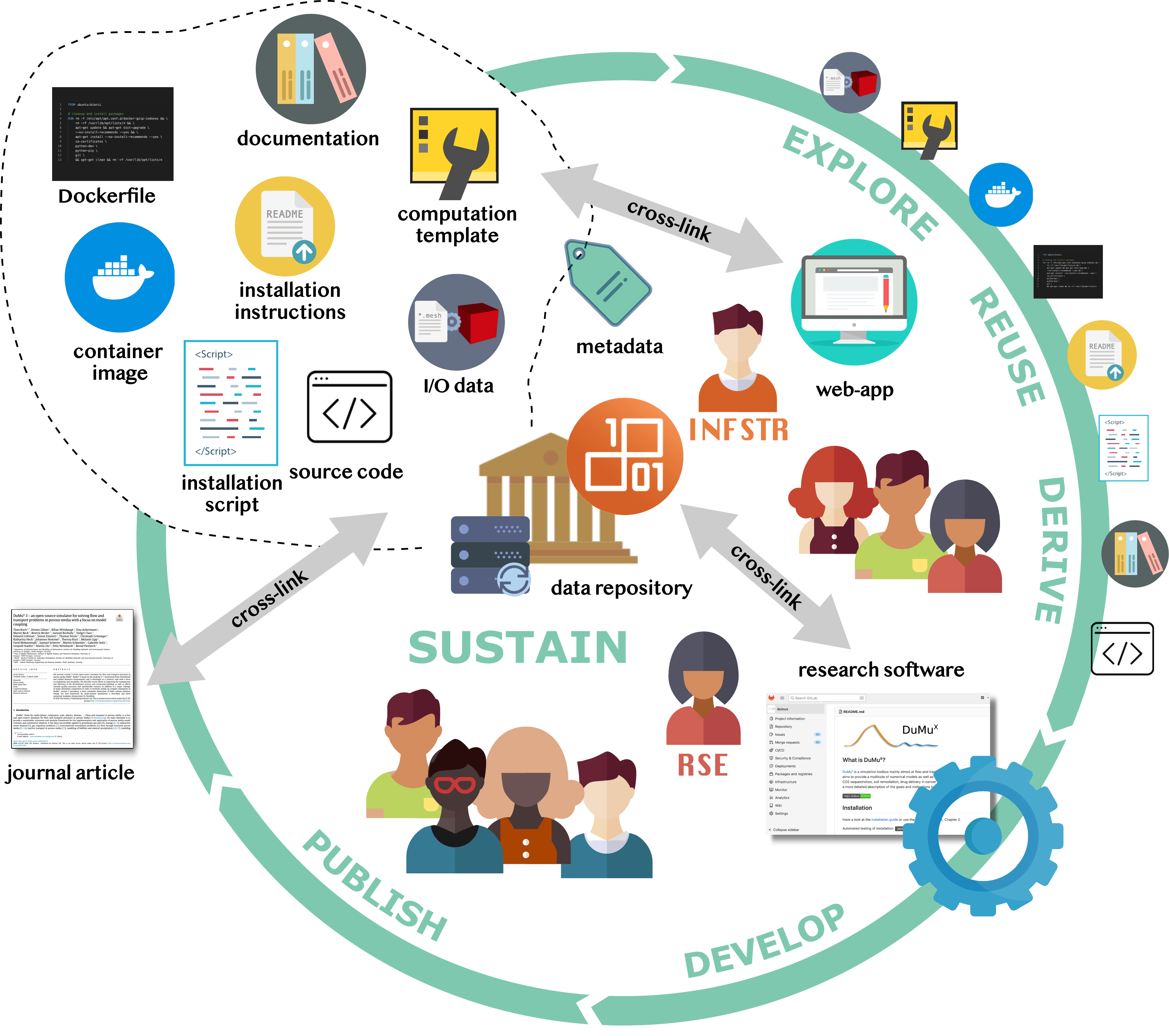}
    \caption{\textbf{Sustainable infrastructure concept for improved
accessibility, reusability, and archival of research
software.} Ideally, research software is developed and scientific knowledge is advanced in the shown sustainable cycle. The researchers publish a snapshot of the research software and research results at the time of publication. These research software artifacts (including the items in the dashed cloud and a possible journal article) constitute a multi-modal representation of the published research that allow others to reproduce and improve upon the original findings. The research software artifacts are published and archived according to the FAIR principles~\cite{Barker2022} for research software in one of multiple data repositories and enhanced by interaction capabilities, for example, a web application for exploration of the results. Reusability (enhanced by interactive archives) and derivability (through accessibility and compatible open software licenses) lead to the extension of existing research software or new software developments. With enhanced or new software, (possibly different) researchers create new results and publications. The sustainable software development and publishing cycle is supported by research software engineers (RSEs) and infrastructure providers (INFSTR), for example, by providing tools for the automation of workflow processes, or by developing domain-specific metadata schemes. All published artifacts are cross-linked through their metadata to offer multiple entry points for finding and accessing the results.}
    \label{fig:sustain_main}
\end{figure}

Previous approaches to archival and reproducibility have often focused on a journal publication and the presented results as a unit~\cite{Fomel2013}, resulting in a single file (\eg\cite{Hinsen2011}) or single archive approaches (\eg\cite{Peng2011}). 
In this work, we want to establish the basis for a sustainable research software lifecycle as shown in Fig.~\ref{fig:sustain_main}, combining best research software development practices~\cite{Jimnez2017,registries2020best} with sustainable and FAIR archival infrastructure~\cite{Druskat2018,hermann2019,Anzt2021}. We allow for a distributed view on data storage, where various research software artifacts may be stored in different (possibly specialized) data repositories, however, always cross-linked and FAIR.

% main propositions
To this end, we propose in Box~\ref{box:maincomp} three major components of sustainable archival, improved accessibility, and reusability of research software and research software artifacts that help move towards this desired state.

\begin{bluebox}{Main components of sustainable archiving}
\label{box:maincomp}

\noindent\textbf{1. Installable, configurable, and extensible research software:} The installation process and its resilience and documentation are an integral part of archived research software artifacts. Research software and related artifacts that are configurable or parameterized, well-documented, human-readable, and editable, are more valuable resources regarding reuse than single-purpose, linear, procedural reproduction recipes. Only free/libre and open-source research software (FLOSS) is considered extensible. \\

\noindent\textbf{2. Multi-modal data representation:} Research software and related artifacts should be archived in a \emph{multi-modal, possibly modular, data representation} (as detailed in Section~\ref{sec:mmdata}) in a way that first access is simple and fast (e.g. practically installation-free \emph{browser-based access}) but instructions for exploration of the software artifacts are available. Setting up a running environment containing the software is assumed to be achievable in several ways, with information of different complexity, and associated with a different amount of work. This is proposed to result in resilience: if the easiest way to reproduce research results or set up a suitable environment for reuse ceases to work at some point in the future, there is still the option to use an additional information layer, additional manual control, and expert knowledge to achieve the task. \\

\noindent\textbf{3. Public open-access archival infrastructure for research software artifacts:} Research software and related artifacts should be archived using \emph{sustainable infrastructure} operating with free open-source software, and with an automated workflow supported by (preferably permanent) staff at the institution of the researcher. The workflow commonly consists of a mixture of domain-agnostic (high-level, application programming interfaces) and domain-specific (low-level, research domain) steps.
%The used infrastructure and software for infrastructure should have a prospect of long-term support at the research institution.
%We suggest that existing infrastructure components originally designed for a different main purpose but with prospective long-term support at the research institution may be repurposed to allow for the workflow as proposed in the following.
This effort usually includes institutional computing facilities \textit{and} library resources to promote FAIR research artifacts~\cite{Wilkinson2016FAIR,Barker2022}. The used infrastructure and software for infrastructure should have a prospect of long-term support at the research institution. To this end, also infrastructure software should be FLOSS (see also Section~\ref{sec:floss}).
\end{bluebox}

All three components are motivated and developed in more detail in the following.
The remainder of this article is split into two main parts. First, Sections~\ref{sec:researach-software}-\ref{sec:components} describe the used concepts, goals, and recommendations for the archival and reuse of software artifacts. The second part, with Sections~\ref{sec:implementation} and \ref{sec:case_study}, illustrates the more abstract first part with examples. In particular, we discuss concrete implementations of the proposed components and experiences within a pilot project at the University of Stuttgart, Germany, which is an ongoing collaborative effort of research software developers, information technology staff, and university library staff.
%In particular, it strengthens the view that a successful largely applicable implementation may require a large number of interacting components (infrastructure and software) and stakeholders rather than a ``single tool to rule them all''.

\section{Research software and sustainability aspects}
\label{sec:researach-software}

We start by briefly introducing the concepts \emph{research software}, \emph{scientific software frameworks}, and \emph{supporting software}, and stress the importance of research software being free and open-source in order to archive it in a sustainable way.

\subsection{Research software, scientific software frameworks, supporting software}
\label{sec:researchsoftware-frameworks}

In this work, we will focus on the development and archiving of \emph{research software}  and \emph{research software artifacts} associated with a research publication, developed by using (domain-agnostic) \emph{supporting software} and recommend developing such software within a (domain-specific) \emph{scientific software framework} facilitating the development and archiving process as outlined in the following. However, we note that scientific software frameworks are usually research software in their own right and, for example, their software design can be a research result worth publishing.

As defined by Hettrick~\cite{Hettrick:2016:RSS}, ``\emph{research software} (as opposed to simply software) is software that is developed within academia and used for the purposes of research: to generate, process, and analyze results. This includes a broad range of software, from highly developed packages with significant user bases to short (tens of lines of code) programs written by researchers for their own use''. (See also \citep{Sochat2022} for a nuanced discussion of a definition.) In particular, in the case of complex programs in terms of software dependencies, archival and reuse requires reproducing a matching environment and is a real challenge.

In scientific computing and engineering (a field several of the authors are affiliated with), the development of scientific software frameworks has been crucial in advancing the field and making the developed technology available to a wider (academic and non-academic) audience\footnote{%
Some well-known influential frameworks in the field of grid-based solutions of partial differential equations are, for example, FEniCS~\citep{Fenics2012}, DUNE~\citep{Dune2008,Dune2021}, deal.ii~\citep{dealii2019design}, PETSc~\citep{petsc-web-page}.
}. (The same is true for many other fields and we cannot provide a comprehensive account.)

With the term \textit{scientific software framework}, we here refer specifically to research software that provides domain-specific abstractions, interfaces, tools, or workflow specifications, meaning: a software ecosystem that facilitates writing application research software for a research project. A framework is typically used pervasively in a researcher's code for the benefit of developing complex flexible applications with relatively little added coding effort. Particularly, frameworks often facilitate the use of well-established third-party software libraries for individual tasks through abstract, domain-specific programming interfaces.

Research software frameworks often contain example applications (or at least tests) that can be used as blueprints from which to start developing. By relying to a large extent on quality-assured code that is tested by both the framework developers and the user base, one reduces the risk of bugs. Less code has to be written and tested, hence more focus can be put on the actual research. Less code also implies less documentation effort. Moreover, the application code is easier to understand for peers familiar with the framework. Finally, and particularly important in the context of sustainable archiving, research software frameworks have the possibility to centrally provide (domain-specific) tools facilitating the archiving of all research code developed within the framework.

However, as a consequence of reusing trustworthy, high-quality software components, a researcher's application has a multitude of software dependencies often requiring to be compiled or installed together (and with specific version and platform requirements) to arrive at a productive environment in which the program can be run and explored, and in which results can be reproduced. Therefore, it is essential for the reproducibility of computer-based research and the reuse of research software that the source code is archived together with comprehensive installation instructions and a description of the computation environment~\cite{Sandve2013}.

Lastly, we distinguish \emph{research software} from \emph{supporting software}. According to \citet{Druskat2018}, the latter ``covers software that is used to create research software, including everything from editors and integrated development environments to version control system platforms''. Supporting software is usually domain-agnostic. It is mostly excluded from the archived software artifacts. In fact, the archived software artifacts should be as independent as possible from the development tools. A notable exception is workflow tools. Supporting software can play an essential role during the automated preparation of software artifacts for archival.

% Emphasize difference research software and software framework in terms of archival

\subsection{Free and open-source (research) software}
\label{sec:floss}

When we refer to research software here, we always mean free/libre, open-source, and redistributable software (FLOSS).
Arguments for FLOSS in a research context and common misconceptions have been discussed, for instance, in \cite{Fortunato2021}.
Adopting a FLOSS license~\cite{Stodden2009,Fortunato2021} (in contrast to a proprietary software license) implies that the software is modifiable (derivable) and redistributable which can be argued to be an \textit{essential prerequisite} for a sustainable research software
landscape~\cite{ChengalurSmith2010,Albers2017,Wilkinson2017,BrownBlog,Jimnez2017,Anzt2021}.
Moreover, it can be argued that out of ethical responsibility, research software that is developed
with the help of public funding should be publicly available~\citep{Anzt2021}.
Besides the software used specifically for research activities, employing FLOSS products is equally important for all infrastructure components~\cite{Albers2017}. When support for or provision of proprietary software ends (\eg due to planned obsolescence), the infrastructure provider is forced to switch to a different product. A bankruptcy of the supplier may lead to the complete loss of knowledge about their proprietary software~\cite{Albers2017}. Moreover, even small extensions of proprietary software will usually involve additional payments. None of these disadvantages exist for FLOSS products.

\subsection{Research software sustainability}
\label{ssec:rs_sustainability}
The goal of the archival process is to increase the sustainability of research software artifacts to the highest degree possible. This is particularly important if the research software is an integral part of the research findings. Wilkinson et al.~\cite{Wilkinson2017} define a hierarchy of sustainability goals
for software partly based on articles by Brown~\cite{BrownBlog} and Hong~\cite{Hong2013}.

\begin{figure}[!htbp]
    \centering
    \includegraphics[width=1.0\textwidth]{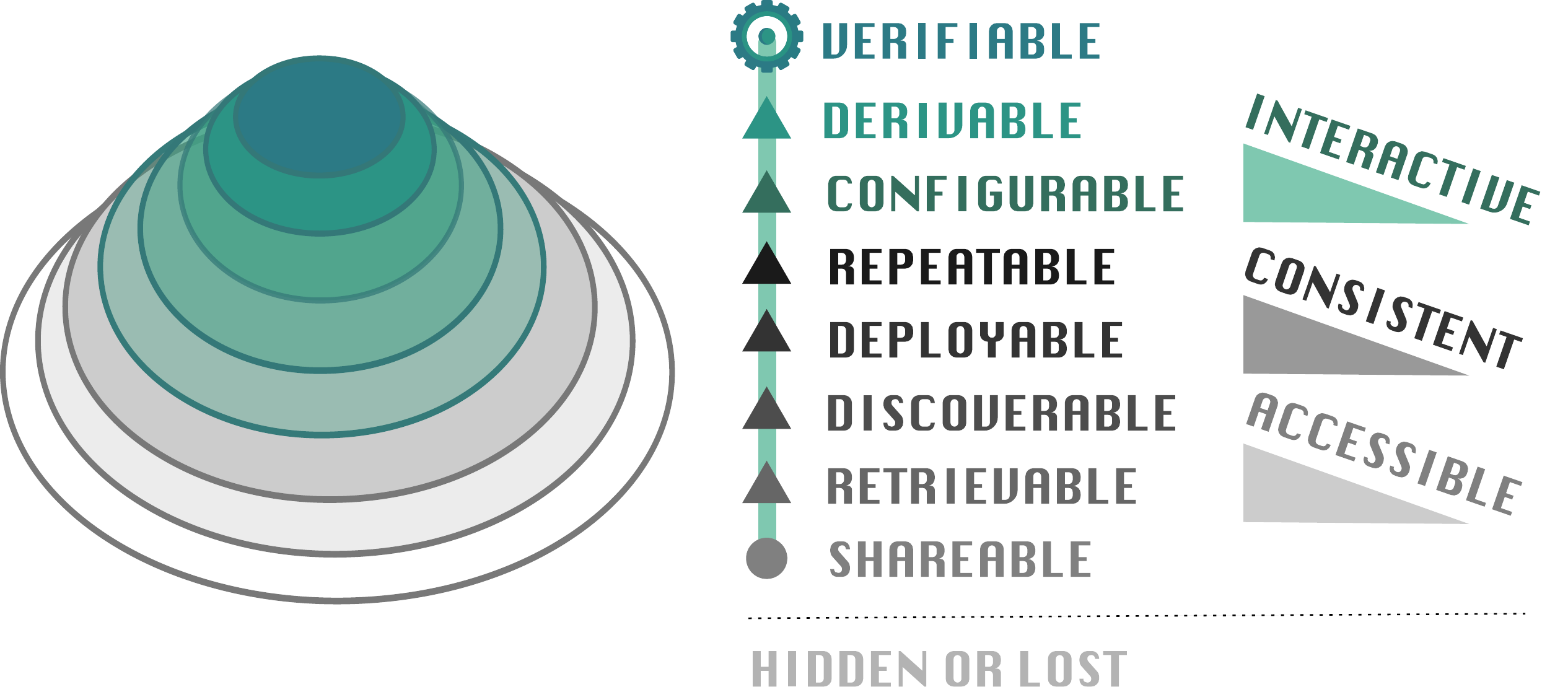}
    \caption{\textbf{The ladder of research software sustainability.} Modified after \citet{Wilkinson2017} (license: CC BY 4.0).}
    \label{fig:sustain_software}
\end{figure}

This hierarchy is illustrated in Fig.~\ref{fig:sustain_software}: being packageable and hence distributable is the minimum requirement for a software artifact to be, in principle, ``shareable'' with the scientific community. ``Retrievable'' artifacts can be downloaded, but practical access may be obscured (hidden location or restricted access permissions). Artifacts become ``discoverable'', and in particular human- and machine-findable, when they are publicly hosted in an indexed repository that is both human- and machine-searchable. ``Deployable'' artifacts can be instantiated and executed, but only ``repeatable'' artifacts produce consistent results over time and on different machines (for instance, by encapsulation in a suitable containerized environment). However, research software cannot be sustained without the following properties: ``Configurable'' software can be applied in new scenarios, while only ``derivable'' software allows for future development by adaption and modification of the source code (compatible licenses implied). Finally, ``verifiable'' software artifacts are correct and their correctness can be tested, for instance, against archived reference data that was produced with the software at the time of a scientific publication. Derivability is thereby essential to verify software in a relevant context not explored in the original scientific contribution.
The sustainability of research software artifacts is ultimately tied to and enabled by a sustainable interactive archival infrastructure~\cite{Wilkinson2017}.

Wilkinson et al.~\cite{Wilkinson2017} discussed existing archival tools in the context of enabling these goals and concluded that in particular interactivity---the ability of archived software artifacts to be reused and modified in new ways---was not yet satisfactorily addressed by most tools at the time. One of the mentioned examples of a highly-interactive archival platform is CodeOcean~\cite{clyburne2019computational}. However, to the best of our knowledge, it is itself not open-source and therefore also excludes the possibility for self-hosting, customization, and extension by the scientific community.
In a sustainable workflow as depicted in Fig.~\ref{fig:sustain_main}, research software and scientific knowledge are in a continuous state of transition where distinct realizations in time are marked by archived research artifacts. To sustain the cycle, research artifacts have to be explorable, reusable, and derivable to ultimately contribute to the development of new or enhanced software.

\section{Community effort and stakeholders}
\label{ssec:roles}

Establishing a sustainability infrastructure is an effort of the scientific community at large rather than an effort of individual researchers, cf.~Fig.~\ref{fig:sustain_main}.
In Box~\ref{box:roles}, we introduce the main roles in the context of sustainable infrastructure and how individuals in various roles interact with archived and reusable research software data artifacts and computation-based scientific results, as well as with each other. The roles introduce a social and a functional setting and suggest some expectations of the different stakeholders towards the infrastructure and the workflow.

\begin{bluebox}{Roles and stakeholders}
\label{box:roles}

\noindent\textbf{Researcher:} The primary interest of researchers lies in publishing new research findings. They would profit from an easy-to-use infrastructure service that enables publishing all research software artifacts, including code, computational environment, input, and result data. Researchers benefit from increased visibility, citability, transparency, and the ability to publish FAIR. An important aspect of practical relevance is that the researchers themselves will likely want to reproduce their own results and reuse their own software months or years later. This is a real challenge. From this perspective, the role coincides with that of a scientific peer.\\

\noindent\textbf{Scientific peer:} The primary target group for a published research object is scientists with a potential interest in the corresponding research findings.
Scientific peers profit from a single access point (including cross-links) to the scientific paper or thesis, associated data, metadata, source code, documentation, and an (interactive) reproducing environment. Whether they are domain experts or not determines the preferred mode of reuse. As most peers won't be interested in all artifacts, the infrastructure must allow for easy extraction of the useful components for, e.g., the comparison of result data artifacts. Web applications lower the entry threshold for using the research software in a well-controlled environment without a complicated installation process and provide peers with the possibility to gain trust in the software artifacts before starting to reuse or derive software artifacts for their own research. Open licenses are essential to peers due to financial constraints,
and---even more importantly---because they allow them to extend the software, to aid their own research projects \emph{without reinventing the wheel}.\\

\noindent\textbf{Research software engineer (RSE):} (We adopt the notion of RSEs
presented in~\citet{baxter2012RSE} as individuals with a background in research, but with
advanced skills in software development and the aspiration to develop tools that can be reused by other researchers.) RSEs benefit from citable research software artifacts they have contributed to. An easily accessible web application may attract more users to the research software framework that the RSE helps develop. Moreover, such applications help build a visible easily-inspected portfolio---an asset for individuals seeking job opportunities in the domain of software engineering and development.\\

\noindent\textbf{Infrastructure provider:} Infrastructure providers have two tasks: they are involved in various research projects to develop suitable infrastructures together with the domain-expert researchers, while at the same time, they ensure that the infrastructure is sustainably operated and built to last.
This includes, above all, maintaining the infrastructure such that published research results and software artifacts can be reused in the long term. Moreover, they enable the interaction of the users with the infrastructure and support researchers. They make sure the infrastructure is modular to allow for adapting or exchanging individual components for future needs. By refining metadata schemes together with researchers and RSEs, they ensure that the archived artifacts remain FAIR.
In collaboration with RSEs, infrastructure providers can develop tools to automate the interaction between researchers, research software, and the infrastructure, thereby simplifying both the publication process and access to research artifacts.
\\

\noindent\textbf{Non-academic user:} Non-academic users may be attracted by public outreach activities or may have a business interest in cutting-edge technology. They might find it difficult to navigate through the latest scientific developments. Therefore, they need easily searchable indices based on accurate metadata. Moreover, they profit in particular from installation-free web applications that do not require extensive domain knowledge to explore the published software artifacts.
\end{bluebox}

Apart from the stakeholders with specific roles, there are additional stakeholders concerned with sustainable infrastructure and publishing FAIR research software artifacts. An detailed account of stakeholders and their relations in the context of sustainable research software environments has been compiled in~\cite{Anzt2021} and \cite{Druskat2018}. Research institutions profit from increased visibility. Synergy effects in joint software development and when optimizing workflows for a large group of researchers at once reduce software development costs. Moreover, reusable software may create opportunities for new research through the extension of existing research software in new directions. Funding bodies likewise profit from less money spent on the redevelopment of software that already exists. Therefore, they may also have an interest in providing incentives for sustainable research software development and the development of research software frameworks. Society at large may profit from more efficient use of taxpayers' money and faster progress in scientific advances due to higher resource efficiency regarding software development (ideal outcome).

\section{Multi-modal representation of research software data}
\label{sec:mmdata}

``Paper[s] of the future'' \citep{Sopinka2020,Goodman2021} are envisioned as multi-media experiences. However, even without heavy use of video, audio, or interactive figures, research software artifacts require multi-modal data representation and cross-linking capabilities. In particular, high-level artifacts (\eg container images) may require little effort for reuse even by non-experts. Nevertheless, the low-level artifacts and procedures used to create high-level artifacts also have to be accessible. In particular, this is the case if a high-level component ceases to work, or users want to test and expand research software artifacts in new environments. This ensures the resilience of the archived dataset. Moreover, different usage scenarios are enabled that target particular roles (Section~\ref{ssec:roles}).

In Box~\ref{box:artifacts}, we describe with \textbf{[A1-4]} the main research software artifacts we have identified and which together enhance reusability in the long term. 
While the source code and associated data, instructions, and documentation may be enough in some cases to ensure reproducibility and derivability (compatible licenses implied), enabling technical reproducibility is not easily verifiable. Practical reproducibility, in particular by non-experts, is unlikely. Practical reproducibility is improved by additionally providing artifacts \textbf{[A5-8]}.

\begin{bluebox}{Research software artifacts}
\label{box:artifacts}

\noindent\textbf{[A1] The source code} is the most central aspect of a research software publication. It contains the code that was required to be implemented by the researchers
in order to produce the data, figures, etc.\, for instance, discussed in a
scientific publication. This also includes configuration files referring to input data or post-processing scripts. In-code documentation is necessary to convey the meaning and literal programming helps readability. However, such research software source code most commonly relies on additional dependencies (such as external libraries or environments such as a Python interpreter).\\

\noindent\textbf{[A2] Installation and execution instructions} are 
necessary (usually even by skilled domain experts) to enable others to (compile and) run the code on their own system. It typically comes in form of a \texttt{README} or a structured text document.
The software dependencies are referenced in their exact versions used. Tested platforms, compiler versions, and hardware requirements are documented.
This is the most basic form of specifying the necessary steps to arrive at functional programs.
Ideally, users are instructed on how to verify that the installed software functions correctly.\\

\noindent\textbf{[A3] Documentation and other publications} describe the intent and concepts down to the function of individual modules of the software in more detail. This can be a journal publication, a user/developer manual, blog posts, etc. Moreover, it may include instructions on how to contribute to the research software or contact developers. Its purpose is to increase trust in and knowledge about the reusability of research software allowing for sustainable and continuous development beyond the published artifact.
To increase citability and cross-referencing, these documents are also published, archived, and made accessible through persistent identifiers.\\

\noindent\textbf{[A4] Computational results, input, and auxiliary data} can be published separately to simplify their reuse independent of the software. Input and auxiliary data are usually needed to (re-)produce results with the software artifacts, \eg for verification. Examples of data reuse independent of the associated research software include performing result-data comparisons against the results obtained with other software or experiments, or further data analysis to derive novel insights. A completely different kind of auxiliary data is detailed information on provenance and the software development processes as typically available in open transparent development platforms (\eg GitHub, GitLab, etc.) in the form of version control history, issue trackers, developer discussion, help desks or mailing lists.\\

\noindent\textbf{[A5] Automated installation and execution} ensures that no manual steps are required to prepare the software environment, executable programs, and the execution of computational pipelines. This may be realized by providing \textit{installation scripts} automating and describing the steps to an executable program, and \textit{pipeline/workflow scripts} automating and describing pre-processing of input data, manipulation of the processed data, and post-processing analyses.\\

\noindent\textbf{[A6] Container recipes} contain the complete set of instructions (including package manager commands) required to arrive at an environment that is suitable to run the considered research application (see Section~\ref{ssec:container-applications}). Container recipes build on top of version-pinned base
images, %(\eg \texttt{ubuntu:20.04} instead of \texttt{ubuntu:latest}),
install version-pinned dependencies (e.g. \texttt{Python 3.9.1}) and run the installation script (reuse).
Even without a resulting container image in mind, such recipes (e.g. \texttt{Dockerfile}s) provide comprehensive both machine-readable and human-readable documentation of the full installation process.\\

\noindent\textbf{[A7] Prebuilt container images} (based on the developed recipes) allow to spin up containerized environments (containers). This provides three further benefits: The computational pipeline can be triggered and the software can be explored in a reproducible (and possibly interactive) environment with minimal effort. Peers can skip the potentially compute-intensive installation process and they do not have to build container images themselves. Finally, container images simplify deployment in cloud services, \eg to be used as the basis of a web application.\\

\noindent\textbf{[A8] Interactive web applications} allow access to the research software's functionality without installation. They can be designed anywhere between a simple graphical user interface for configuration, execution, and visualization\footnote{Examples for FLOSS libraries facilitating the creation of web applications are \texttt{Shiny} for R~\citeW{shinyr} and \texttt{streamlit} for Python~\citeW{streamlit}.} (suitable in particular for non-experts and in first-contact scenarios, cf. Fig.~\ref{fig:viplab-frontend} and Section~\ref{ssec:interactive-reuse}) and an interactive workbench environment\footnote{In its notion used here, an interactive workbench environment differs from a full integrative development environment (IDE) in the lack of debugging tools and a fixed rather than configurable build automation process. Moreover, it is a web application rather than a desktop application. One example is the proprietary platform CodeOcean~\cite{clyburne2019computational}.} with the possibility to modify source code and reevaluate the outcome (exploration and derivation for domain experts). Such web services often rely on containerized applications, for which the published image can be reused. Infrastructure providers at academic institutions facilitate the long-term availability of interactive environments.
\end{bluebox}

Finally, \textit{cross-linking} all research software artifacts belonging to the same research object
is essential when they are stored in different locations. For example, the journal, software, and data publications as well as the source code repository in which the code was (and possibly still is) developed are all cross-linked among each other~\cite{Maric:2022:RSE}.
It can be argued that it might be ideal to have a single dataset that represents an entry point to a research object and links to all resources stored in other locations via persistent identifiers. However, the unit of a research object often has no clearly defined boundaries and there may be both hierarchical and distributed elements to its logical structure.
Accurate and descriptive metadata together with suitable domain-specific ontologies allow mapping the interaction and interdependency of research software artifacts in distributed archival environments. To guarantee compatibility with other repositories, standardized schemes have to be employed (FAIR principles). Cross-links are findable and traceable by both humans and machines.

\section{Recommendations for components and infrastructure}
\label{sec:components}

The notion of multi-modal data---as described in the previous section---poses certain requirements on both the software itself and on the infrastructure on which it is published. From our (ongoing) experience in connecting different components and infrastructure to improve accessibility, reusability, and archivability (see Section~\ref{sec:implementation}), the following sections compile concrete recommendations on how to meet these requirements for research software in Section~\ref{ssec:research-software}, containerization in Section~\ref{ssec:container-applications}, interactive web applications in Section~\ref{ssec:interactive-reuse}, and multi-modal data repositories in Section~\ref{ssec:multi-modal-data-repo}.

\subsection{Research software and research software development}
\label{ssec:research-software}

To be of significant value to the scientific community, we think that research software publications should contain ``derivable''\footnote{For descriptions of the terms in double quotes, see Section~\ref{ssec:rs_sustainability} and Fig.~\ref{fig:sustain_software}.} artifacts that enable peers to adapt the software for their own research. This requires the software to be installable and documented. Installability of published research software remains a major concern~\citep{Mangul2019b}, which highlights the importance that research software should be developed in a portable way, for instance, by relying on cross-platform build systems~\citep{Maric:2022:RSE}.

Addressing the issue of installability, we recommend to follow \textbf{[RS1-4]} in Box~\ref{box:recommendation-software}. For a software publication to be ``verifiable'', it must provide executable artifacts that are able to reproduce the results discussed in a scientific publication. Such results are often the product of multiple computational steps, typically involving pre-processing of input data, manipulation of the processed data, and post-processing analyses. For users to be able to verify published software artifacts, reference results and the configuration and input data used to produce them must be available~\cite{Hutson2022}. To achieve this with minimal overhead, we recommend \textbf{[RS5-7]} in Box~\ref{box:recommendation-software}.

\begin{bluebox}{Recommendations for research software}
\label{box:recommendation-software}

\noindent\textbf{[RS1] Build on existing research software frameworks} and reuse existing software as much as possible. Frameworks often implement quality assurance measures (increases trust) and may provide tools to create portable software environments.\\

\noindent\textbf{[RS2] State all dependencies} of your software including the exact version used in the workflow. %In general, relying on frameworks helps in this regard as they provide documentation on their dependencies and the installation process.
\\

\noindent\textbf{[RS3] Create installation scripts} for all dependencies of your software that cannot be obtained via widely-used\footnote{\sffamily Whether a component is considered widely-used is in general community-specific.} package managers (\eg \texttt{apt}, \texttt{pip}, \texttt{conda}, \texttt{guix}, etc.). This is particularly important in the common situation where a dependency is itself a research software and has to be retrieved in the version specified by a commit hash in a version control system, or when a dependency has to be patched (source code modification).
%Frameworks may provide templates or mechanisms for generating installation scripts for downstream applications.
\\

\noindent\textbf{[RS4] Use persistent identifiers} in your installation scripts to retrieve dependencies. To this end, consider using the Software Heritage archive~\citeW{SH}~\cite{DicosmoSoftwareHeritage2017,DiCosmo:2020:ARS} providing these on a source code repository, file, and single line level.\\

\noindent\textbf{[RS5] Automate computational pipelines} from the beginning of a project. This enables users (and yourself!) to repeat the process without a detailed understanding of the individual steps that have to be performed. Automation can be realized by scripts or by using a portable workflow language~\citep{Diercks2022Preprint}.\\

\noindent\textbf{[RS6] Version-control} code and configuration/input data %\comment[id=as]{Do we really recommend to version control the input data? This can be quite large or (unique) experimental data?} --> we require people to think a bit on their own
required for the workflow (including the workflow description). Use defined versions of your configuration/input data in your workflow description. This approach avoids mismatches between published computational results and software, hence is more likely to yield ``verifiable'' artifacts. Tools like \textit{Data Version Control} \citeW{dvcontrol} can provide additional help to map each computational result to specific versions of code and input data.\\

\noindent\textbf{[RS7] Containerize applications} for when they cease to be ``deployable`` (at least without effort) in the constantly evolving software and hardware environments. Containers are more likely to preserve a ``verifiable'' software artifact over longer time scales, and can serve a variety of further purposes (see Section~\ref{sec:mmdata}). Frameworks can again be of help here as they may provide templates or mechanisms to create container images for downstream applications.
\end{bluebox}

\subsection{Containerized programs}
\label{ssec:container-applications}

Containerization can be used for several purposes (e.g. rerun workflows, reproduce workflows, automated testing pipelines, deployable web applications, etc.) and are therefore valuable assets promoting ``Deployable'' and ``Verifiable'' research software (also see Sections~\ref{sec:mmdata}, \ref{ssec:research-software} and \ref{ssec:interactive-reuse}). To increase the sustainability of container images published as research software artifacts, we give recommendations in Box~\ref{box:recommendation-container}.

\begin{bluebox}{Recommendations for containerized programs}
\label{box:recommendation-container}

\noindent\textbf{[CP1] Publish container build recipes} alongside prebuilt container images to make their creation transparent and verifiable. Publishing build recipes makes them ``Derivable'' artifacts suitable for modification and extension.\\

\noindent\textbf{[CP2] Avoid hardware-specific compiler flags} when compiling software in containers to improve portability.\\

\noindent\textbf{[CP3] Use version-pinned base images and packages} in your recipes, so images can still be built at a later point in time and/or on other architectures. For instance, use \texttt{ubuntu:22.10} instead of \texttt{ubuntu:latest} (as the latter points to the most recent version instead of a fixed version). Pin versions consistently: for example, when pinning the \texttt{Python} library \texttt{numpy}, also pin the \texttt{Python} version.\\

\noindent\textbf{[CP4] Keep container images small} by storing large input data and auxiliary files externally (see \textbf{[A4]}) and fetching them via persistent identifiers. Follow technology-specific and domain-specific best practices (e.g. remove package manager caches; see also \citeW{d-docker-bestpra} and  Section~\ref{sec:docker}).\\

\noindent\textbf{[CP5] Generate images compliant with technology-specific standards} such as those developed by the Open Container Initiative (OCI)~\citeW{oci}. (For example, Docker images follow OCI standards.)
\end{bluebox}

The creation of container images requires particular skills and knowledge about container technologies. Research software frameworks may provide templates or automation tools. Researchers may be assisted by an RSE. A number of FLOSS tools aim to help with the containerization of research code\footnote{For example, Dockta~\citeW{Dockta} simplifies the creation of containers for applications written in R, Python, NodeJS, JATS, and Jupyter. Derrick~\citeW{Derrick}, repo2docker~\citep{Repo2Docker2018}\citeW{Repo2Docker}, Whales~\citeW{Whales}, containerit~\citep{containerit2019}\citeW{containerit}, and Source-To-Image (S2I)~\citep{s2i}\citeW{s2iurl} are similar examples of such tools.}.

Another benefit of containerized research software is its use in continuous integration and delivery (CI/CD) pipelines used for automation in the software development (build and test) and deployment (release and operate) cycle and software quality assurance. There, containers ensure reproducible environments.

\subsection{Interactive reuse of software and data and browser-based access}
\label{ssec:interactive-reuse}

Web applications that run the research software remotely have the advantage that users do not have to meet any hardware or software requirements, but only need access to a browser and the internet. This facilitates achieving practical reproducibility, and the research software has to be maintained in a ``deployable'' state only on the remote infrastructure.

The main goal of the web application is usually to repeat (some or all of) the computations related to a publication. If the application includes automated checks to published result data, the research artifacts become ``verifiable''. In principle, (possibly headless) web applications with an application programming interface (API) could be used for automating the verification of research software. The infrastructure could trigger the verification step regularly and automatically, making it transparent to the users if reproduction can be achieved. This would increase trust in the research software regardless of its actual usage statistics.

Another important goal (in the context of sustainability) is the exploration of the software and data beyond the published result in addition to reproducing the published result. ``Derivable'' software is promoted whenever the web application allows for (partial or full) modification of the original source files and parameter settings. To this end, (as mentioned in \textbf{[A8]},) web applications may be anything between a simple graphical user interface (GUI) application to configure and launch computations, and an interactive workbench that allows for extensive source code modifications and build and deploy cycles. With the following recommendations, we focus on simple GUI applications. In particular, the goal is to create an interactive user experience ideal for first-contact exploration scenarios \footnote{We assume here that, in practice, the web application might not be the perfect reproduction and reuse environment in all reuse scenarios. A more thorough exploration or extension of the software, as well as in some cases the full reproduction of all research results, may require installing the software locally on the user device, instead. For this purpose, research software datasets must contain research software artifacts of other data modalities, see Section~\ref{sec:mmdata}.} of research software artifacts. To achieve this, we recommend \textbf{[WA1-5]} in Box~\ref{box:recommendation-webapp}. To enable researchers to adapt web applications to their needs, we recommend infrastructure providers to create configurable web application templates, \textbf{[WA6-7]}.

%\comment[id=dg]{We should discuss WA6 and WA7, such that we can mention them here in a similar way as we refer to 1-5}

% I (Dennis) commented the following because in the discussion we alread mention under which circumstances the additional work is reduced. I think also in the use case we talk about this...
%Naturally, creating such applications requires additional work by the researcher. This work can be reduced to choosing the configurable parameter space, if the software is configurable, the workflow is automated, and the application is containerized (see previous sections). We recommend \textbf{[WA6]}.

\begin{bluebox}{Recommendations for (simple) web applications}
\label{box:recommendation-webapp}

\noindent\textbf{[WA1] Choose a meaningful and limited set of parameters} to be configurable in the browser-based interface. It should provide an intuitive way to explore the software, possibly also for interested users outside of academia. Bloating it with too many and too obscure options misses the purpose. \\

\noindent\textbf{[WA2] Use the parameters of the publication as defaults} such that the verification of the research software can be done without having to manually extract parameter values from the publication.\\

\noindent\textbf{[WA3] Prefer sliders or radio buttons} over free input in order to avoid mistakes due to faulty or incompatible input and thereby improve the user experience. Where possible, define ranges of validity for the input parameters and let the infrastructure validate the values.\\

\noindent\textbf{[WA4] Hide implementation details and boilerplate code} when exposing source code files to the user for modification. Ideally, only those parts of the code that are required to understand/modify the key algorithms of the research are visible to the user. \\

\noindent\textbf{[WA5] Optimize the web application and the offered reuse scenarios for reactivity} to enhance interactivity. For longer-running applications deliver intermediate results and visualizations. \\

\noindent\textbf{[WA6] Provide configurable web application templates}
%\comment[id=as]{This strongly depends on the specifc framework used?! Should we maybe only recommend to put all stuff needed to create the webapp as artifact in the repo?}
such that 
\begin{enumerate*}[label=(\alph*)]
\item creating new applications is a manageable effort,
\item researchers can control application design and functionality. 
\end{enumerate*}
%Application templates should provide both fixed components and optional, configurable components.
\\

\noindent\textbf{[WA7] Make web application configurable by a file}, so that the description of web applications is transparent and archivable. Ideally, a simple open file format is used (e.g. JSON). Researchers should be enabled to publish web application configuration files in the dataset alongside code and deployment artifacts, cf.~\textbf{[A1,A5-7]}.
\end{bluebox}

%Beyond, the interface should be able to simplify the configuration of the research software to avoid mistakes of the user or to streamline the interaction. For that, GUI components such as input fields, sliders or checkboxes can be useful as well as the possibility to hide boiler plate code that is needed to run the software and reproduce results, but is not needed to understand the algorithms / research work. This also reduces the complexity so that users are able to focus on more important parts of the publication. 

%The browser-based interface should also be replaceable, such that it is possible to realize different kinds of visual appearance according to the specific needs of the research software. This can be realized by a modular structure of the application in which individual components are exchangeable. As a result of this, the sustainability and maintainability of this component is increased. 

Remote code execution raises security concerns and, depending on the use case and target group, scalability may be an issue from the infrastructure point of view. To run applications, a computing infrastructure has to be established in collaboration with institutional infrastructure providers or cloud service providers. 
For security reasons, research software and in particular software configured with user input has to be run in sandboxed environments. Any user input collected in the web application or data fetched via persistent identifiers (\textbf{[CP4]}) must be validated prior to program execution. Ideally, research software container images are designed such that executing the program does not require an active internet connection. Disabling specific CPU features or constraining the amount of CPU and memory resources that can be acquired by the applications may further prevent misuse.

\subsection{Multi-modal software data repository for improved reuse}
\label{ssec:multi-modal-data-repo}

In general, research data and research software artifacts should be published and archived according to the FAIR principles~\cite{Wilkinson2016FAIR,Barker2022}. For multi-modal data in particular, we recommend \textbf{[MR1-6]} in Box~\ref{box:datarepo}.

\begin{bluebox}{Recommendations for multi-modal data repositories}
\label{box:datarepo}

\noindent\textbf{[MR1] Assign persistent identifiers} on both the dataset level and for individual files to ensure citability and promote the reuse of individual assets. \\

\noindent\textbf{[MR2] Allow multiple file formats.} A multi-modal repository should not be restricted to certain file types, but allow the creation of coherent datasets with multi-modal artifacts. Data in ancillary repositories specialized in the FAIR storage of certain data types (e.g. git repositories) can be cross-linked. \\

\noindent\textbf{[MR3] Allow for and make use of domain-specific metadata} and make it possible to add new metadata schemes in a modular way using standardized format descriptions. Metadata schemes should be shared~\cite{registries2020best}. Use standardized formats such as DDI~\citeW{ddi}, 
%\as{i could not find a real reference for DDI - Sibylle, maybe you have one?}
DataCite~\cite{datacite-standard}, EngMeta~\cite{darus-500_2019}, and CodeMeta~\cite{codemeta-standard}.\\

\noindent\textbf{[MR4] Allow to choose different licenses} for individual assets. \\

\noindent\textbf{[MR5] Implement a review process} for newly added datasets to assure quality (metadata, licenses, practical reproducibility).\\
%Ideally, this process is automated. To ensure correctness, the dataset needs to provide an automated way of determining reproducibility. Manual review and data curation may be required in some instances. \\

\noindent\textbf{[MR6] Enable interactive exploration} with domain-agnostic technology but domain-specific configuration (by the dataset creator); see Section~\ref{ssec:interactive-reuse}. The type of exploration tool can depend on the file type and additional metadata.
\end{bluebox}

%We recommend to first think about the repository in which the data can be published. Usually, the research software is managed within a git repository, but a repository for really publishing the data need to address more to be FAIR.  The repository should assign PIDs not only on a dataset level, but also on all files.

%It should be possible to link to other sources and the repository should not have no file format restrictions.

%The repository should provide the possibility to describe datasets with different domain-specific metadata and to choose an appropriate licence. Also, desirable would be a license checker that checks if there are dependencies in the software to be published.

%For our proposed practically reproducible, the repository should provide the opportunity to try out what has been published.
%It would also be desirable to have some kind of feedback function that shows how often and successfully the data has already been run.

%In order to implement the FAIR principles, we recommend to establish an external review process.
%This can comprise checking of the metadata, READMEs, licences and file naming.

Efficiency in the publishing process requires automation. It is necessary to have supporting infrastructure staff working towards providing suitable metadata, and appropriate linking of keywords to ontologies~\cite{UscholdOntologies2015}. Newly developed research software and research fields may require new domain-specific metadata schemes to ensure FAIR artifacts.

\section{Implementation example}
\label{sec:implementation}

This section describes concrete technical realizations of the proposed components and illustrates some issues experienced in their implementation.

Specifically, we describe our effort in a pilot project over the course of four years in realizing the proposed components by reusing, developing, and connecting existing infrastructure at the University of Stuttgart. We first identified suitable components that have the prospect of long-term institutional support. All selected components have existing applications and usage patterns and have not been designed specifically for this study. Based on the existing properties of these components, we describe the adopted modifications to integrate the components into the proposed sustainable infrastructure concept.

\subsection{Research software framework: example \dumux}
\label{ssec:dumux}

This section describes the research software \textit{framework} \dumux. When it comes to archiving, it is important to note that in this works, we focus on archiving research software (associated with a research publication) developed within the framework rather than archiving the framework software itself. We recommend with \textbf{[RS1]} to make use of frameworks in the development of research software and make use of tools provided by the framework facilitating the archiving process.

As an example of such a framework, we chose the open-source porous-media simulation framework \dumux~\citep{FlemischetalDumux2011,KochetalDumux2021}~\footnote{Several authors are active developers of \dumux.}. (The choice of framework is often domain-specific. Researchers from different fields will certainly choose a different framework.) The \dumux project has several properties that make it suitable (but by no means unique!) as a demonstrating example:
\begin{enumerate}
    \item High complexity as a modular \cpp project, based on the Distributed Unified Numerics Environment (DUNE)~\citep{Dune2008,Dune2021}\footnote{A long-lived (20 years) internationally recognized open-source software framework for grid-based scientific computing.} and several dependencies with non-trivial installation process (requiring configuration and compilation with their own software dependencies)
    \item The developing community satisfies some precursors for sustainability \citep{ChengalurSmith2010,Coelho2017} and follows many best practices for research software development~\citep{Jimnez2017}, for example:
    \begin{enumerate}
        \item Version control (Git), contribution guide~\citeW{1-dumux-contrib}
        \item Transparent, accessible, open development process on a GitLab instance~\citeW{0-dumux}
        \item FLOSS license (GNU GPL 3~\citeW{gpl3} or later)
        \item Active developer and user community with mailing list~\citeW{2-dumux-ml}, 
        \item Extensive documentation~\citeW{3-dumux-docs} (source code with Doxygen, handbook, open course material for self-study~\citeW{a5-dumux-course})
        \item Code of conduct~\citeW{4-dumux-coc}
        \item Regular releases, backwards-compatibility policy between minor version updates~\citeW{5-dumux-reltag}
        \item Automated continuous integration pipelines (CI/CD) using containerized environments~\citeW{6-dumux-pls}      
        \item Extensive test suite with more than $500$ tests (unit tests, integration tests, system tests, verification benchmarks), test coverage report~\citeW{7-dumux-coverage}, and dedicated testing toolset for the use in downstream modules~\citep{Kempf2017}
    \end{enumerate}
    \item Mature code-base with more than ten years of active development~\citeW{8-dumux-openhub}, used internationally in multiple research groups~\citeW{9-dumux-pubs}, modern methods due to research affinity in method development
\end{enumerate}

%\tk{This should be what we changed and what problems we had, not what is good about dumux. The following needs complete rewriting. Less bragging, more specifically what was improved. E.g. scripts, Python interface, CI/CD tests installation process in Docker. Adding Docker to dumux-pub. metadata extraction (although not connected yet).}
%\comment[id=bf]{I added Python bindings, reformulated/expanded the paragraphs on the installation script and dumux-pub and moved them to the Docker section.}
%\tk{Cool makes sense in the Docker section.}
%\tk{I think this part can be shortened significantly.}

In the following, we describe some features and components associated with \dumux related to the suggested infrastructure concept and describe implemented modifications.

Configurability is an important aspect of \dumux for its usage as a research code. This is technically realized by template-based \cpp programming techniques (traits and policies) at compile time and configuration files (\texttt{INI} file format) and a command line argument parser at execution time.
The configurability was enhanced by the introduction of Python bindings based on corresponding developments in the base framework DUNE (starting with version 2.8). Python bindings have been implemented for a subset of the \dumux functionality. Python-based tests have been developed and added to the continuous integration (CI) pipelines~\citeW{dp-python-programs}.

The \dumux source code is hosted, developed, and maintained on a GitLab server run at the
Department of Hydromechanics and Modelling of Hydrosystems (LH2) at
the University of Stuttgart.
Besides the main \dumux source code, the GitLab instance hosts a large number of related projects
such as the \dumux course material~\citeW{a5-dumux-course}, 
\dumux code used for teaching~\citeW{a4-dumux-lecture}, 
\dumux code used in publications~\citeW{a3-dumux-pub}, and more.
Affiliated users can create new projects (close to the code base) which enables more direct contact with developers and cross-linking capabilities to other projects. A code of conduct~\citeW{4-dumux-coc} has been adopted to promote inclusivity in the user and developer community.

The \dumux documentation consisting of multiple resources~\citeW{3-dumux-docs} 
that have been enhanced by adding documented examples that are rendered in GitLab~\citeW{a6-dumux-ex}. These examples have been included in the test suite (CI).

GitLab has built-in support for continuous integration pipelines (CI/CD). Within the effort to increase sustainability, the pipeline repertoire has been extended with cross-project pipelines to verify that changes to \dumux do not break important downstream projects that depend on it. 
The developed pipeline scripts can also be reused in modules developing research applications in the context of a journal publication.
Finally, we added pipelines that automatically create and deploy Docker images for different test environments into the container registry of the GitLab instance.
\dumux in containerized applications in the context of reproducible environments is discussed in the following section. 

\subsection{Containerized programs: example Docker for \dumux-based applications}
\label{sec:docker}

\begin{figure*}[htbp]
  \centering
  \includegraphics[width=1.0\textwidth]{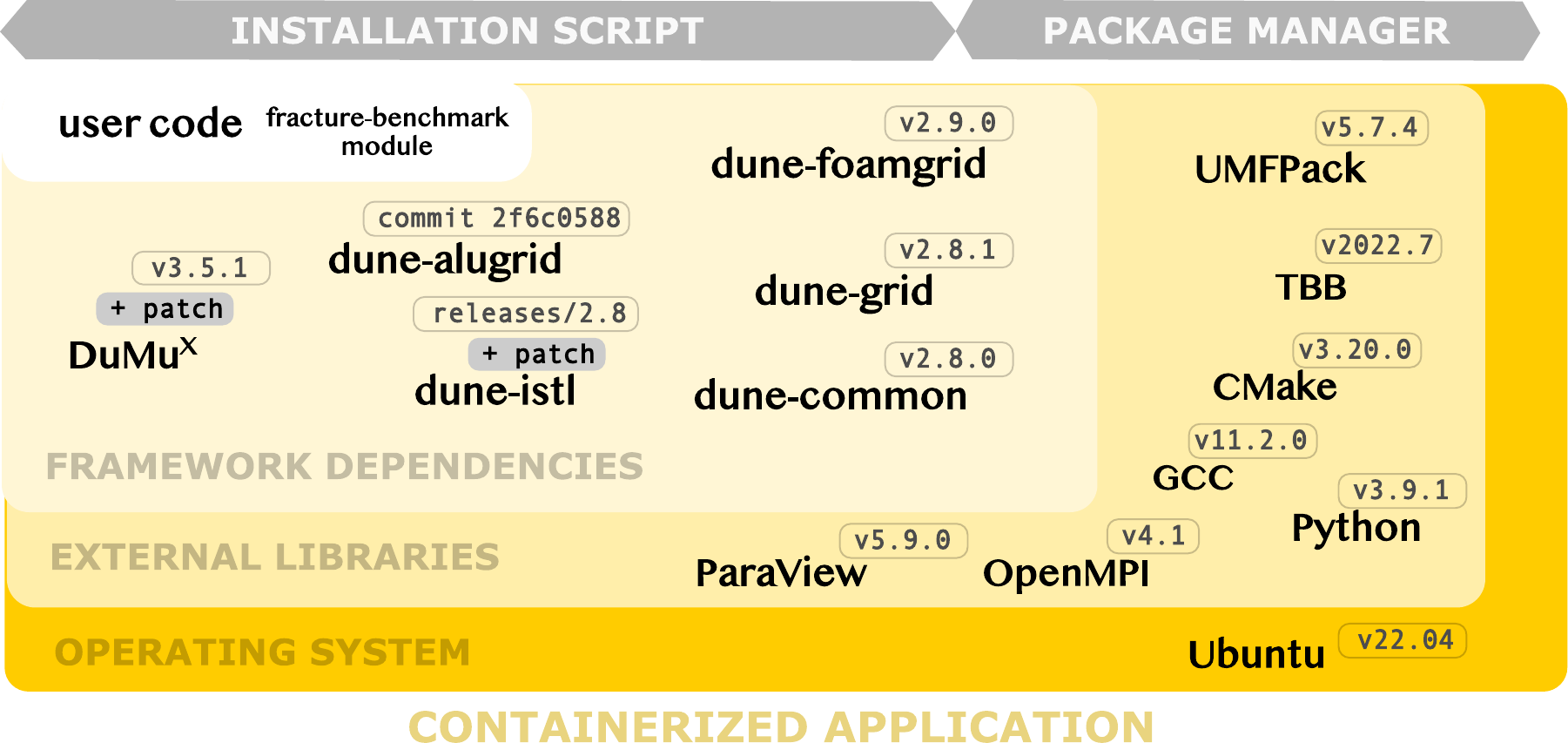}
  \caption{\textbf{Exemplary software environment for a user application building on the \dumux/DUNE framework.} We distinguish software dependencies on the framework level, external libraries, and basic system libraries of the operating system. All dependencies are versioned. However, framework dependencies might not be on released versions. They may be installed from source, the used version (revision) might be only given by a commit hash value and there might be local modifications (patches). We distinguish these packages from such dependencies that can be installed in the required versions via package managers. Unstable versions or software not contained in standard package managers are installed by an installation script in our approach.}
  \label{fig:environment}
\end{figure*}

% general considerations with Docker
We have chosen Docker~\citeW{webdocker}, the most prominent environment at the time\footnote{Another popular container environment in scientific computing is Apptainer (Singularity) which maintains an interoperability layer to Docker}, to create containers, container recipes (\texttt{Dockerfile}s) and (OCI-compliant) container images. During the scope of this project, the process of generating Docker images for \dumux applications has been automated by an RSE by means of a combination of template files and Python scripts (published as part of the \dumux software). 

A necessary component for automating the creation of Docker images for \dumux applications was to capture a user's DUNE environment and document the instructions required to recreate this environment.

As shown in Fig.~\ref{fig:environment}, \dumux applications depend on \dumux itself and a number of DUNE modules\footnote{In the context of the DUNE environment, the \dumux software is itself also considered a DUNE module.}, each a separate software library providing specific functionality. The modules may follow different release cycles and naming conventions. Typical \dumux application developers download the sources of these modules in the versions they require, and configure and build their application locally. Although not usually necessary, they might also have done some local modifications (applied ``patches'') to any of these modules.

With this setup in mind, a Python script~\citeW{a1-dumux-instsc} was developed that determines the versions of all modules an application depends on, including the detection of local modifications. The script then exports the instructions for downloading, possibly patching (in case of local modifications), and configuring all DUNE dependencies into an \textit{installation script}.

Another Python script developed in this project~\citeW{a2-dumux-cdi}, creates a \texttt{Dockerfile} for a \dumux application. The created \texttt{Dockerfile} builds on a version-pinned base image containing the operating system, installs external libraries via package managers, and reuses the application's installation script for setting up the DUNE environment, as illustrated in an exemplary configuration in Fig.~\ref{fig:environment}.

Another development concerns \dumux-based research software artifacts published prior to the mentioned improvements. Since 2015, the \dumux developers at the University of Stuttgart have been running the multi-year pilot project \dumux-Pub which
aims at accompanying each publication (student theses, journal publications) with a corresponding software repository (Git repository) to store source code and essential documentation to reproduce the reported results. At the time of writing, $165$ such modules had been uploaded~\citeW{a3-dumux-pub}.

In an ongoing effort, we retrospectively added \texttt{Dockerfile}s and container images to these repositories with the help of the above-mentioned script~\citeW{a2-dumux-cdi}. Images for 59 modules were successfully created. For some modules, however, Docker images could not be created without significant effort. The main reasons included: missing or imprecise (\eg not version-pinned) listings of dependencies, dependencies on inaccessible code, or missing files essential for installation.

With both the above-mentioned scripts in place, we expect that such errors can be ruled out in the future. By construction, the installation procedure is automatically tested when building the container image before publication. Researchers today can create container images for their \dumux applications with negligible effort, leaving more time to verify that the containerized software indeed reproduces the results to be published. The created installation scripts,  \texttt{Dockerfile}s, and container images can be (and should be) published together with the source code. (A pilot case publishing a specific \dumux-based research software follows in Section~\ref{sec:case_study}.)

\subsection{Browser-based access: example ViPLab}
\label{sec:viplab}

The Virtual Programming Laboratory (ViPLab) was initiated at the University of Stuttgart in 2009.
The initial goal was to create an online tool to teach programming and to integrate that tool into the existing e-learning infrastructure as a plug-in of the ILIAS~\citeW{il} Learning Management System. ViPLab supports programming exercises with different programming languages and software (such as \dumux)~\cite{richter2016viplab,richter2012viplab}.

\begin{figure*}[htbp]
  \centering
  \includegraphics[width=0.9\textwidth]{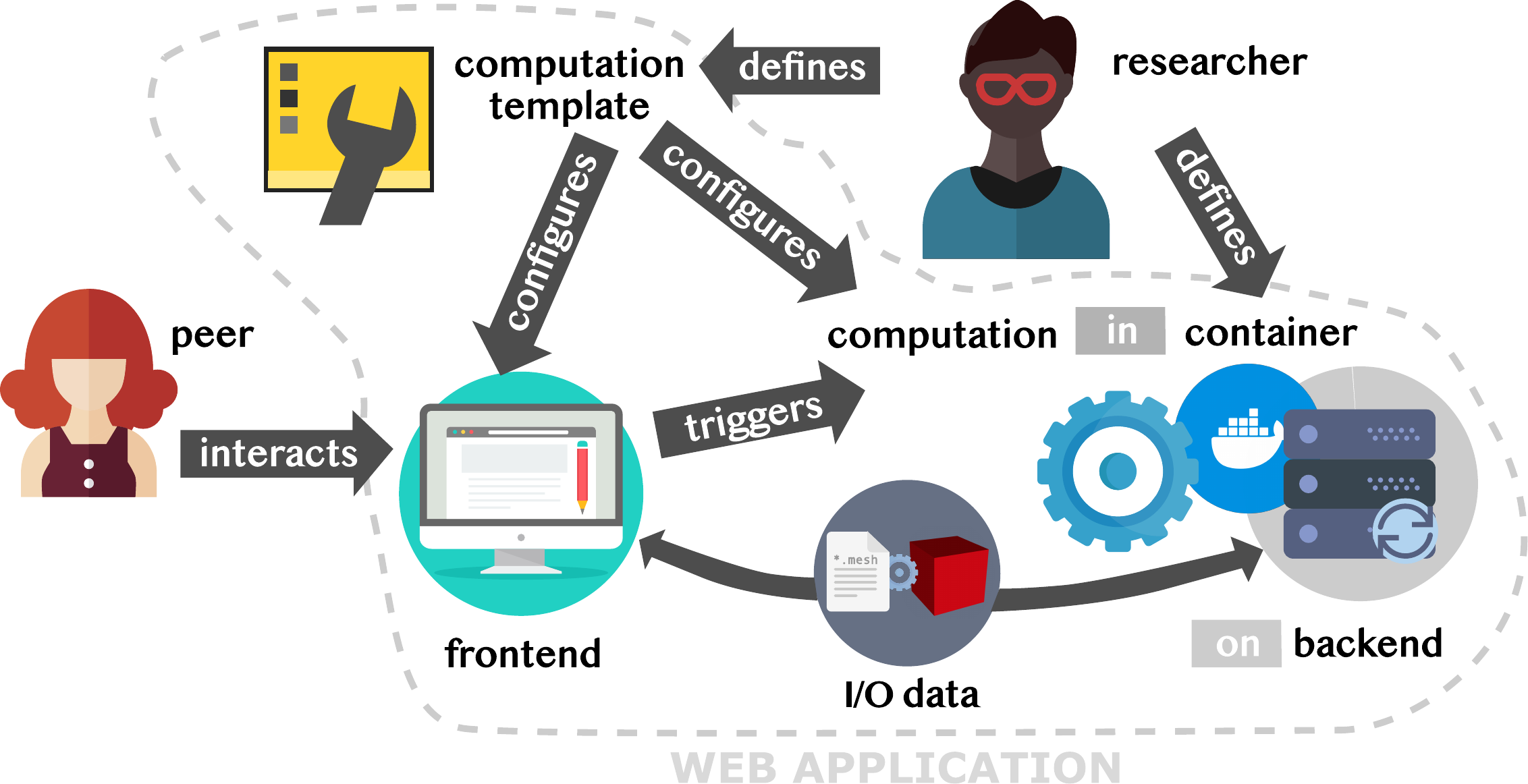}
  \caption{\textbf{Basic components and their relations for a web application designed for exploration and reuse scenarios.} Researchers design a computation template and container image for their application. (In ViPLab, the computation template can be designed through a browser-based GUI application.) The computation template fully configures the application including the GUI of the frontend. Scientific peers use the browser-based GUI in the frontend to configure, run, and reproduce computations and visualize results.}
  \label{fig:webapp}
\end{figure*}

To this end, ViPLab provides a browser-based programming environment that enables teaching instructors to design, and students to work on the programming exercises remotely.
We transferred this concept to the research community: Researchers design a configurable browser-based application for their archived research software artifacts. A scientific peer uses the browser-based application to configure the software. The configured program is executed remotely and resulting artifacts (e.g. output for visualization) are retrieved. The basic components and processes of such a web application are illustrated in Fig.~\ref{fig:webapp}.

The ViPLab environment was deemed a suitable (but again, similar to \dumux in Section~\ref{ssec:dumux}, by no means unique) candidate for the implementation of this concept because its original components already fulfilled most of our recommendations and security requirements, cf. Section~\ref{ssec:interactive-reuse}. Moreover, we anticipated that a software general\footnote{Generalization can also lead to better software abstractions, which we set as one of your goals for the software development.} enough to support teaching and research scenarios has a higher potential for sustainable support through a larger potential user base.
 
From an architecture viewpoint, ViPLab consists of a backend for executing programs and returning any resulting data, a middleware for frontend-backend communication and load balancing computations, and a frontend accessed by the user through a browser. To modernize the code base and support research and development applications, all components were adapted or replaced: backend, middleware, and frontend.

A new backend was implemented allowing for the execution of containerized applications based on (OCI-compliant) container images (see Sections \ref{ssec:container-applications} and \ref{sec:docker}). Thus, ViPLab now supports a wide range of research software independent of software dependencies or programming languages.
The container shares a storage area with the backend file storage system. The storage area is used to exchange files between backend and frontend. These can be both input files sent by the frontend (exchanged before running the program) and output files generated in the backend (exchanged during program execution or after the program has finished).

A new frontend was designed and implemented as a JavaScript-based single-page application that can either be used standalone or embedded in a web page.
The frontend allows configuring a program via a browser-based GUI, triggering the execution of the program, and visualizing or downloading resulting data (\eg by a scientific peer who wants to explore an archived research software application).

Different research software requires differently configured web applications. ViPLab is designed to be highly configurable with some fixed elements and many configurable elements. Designing a new web application means specifying the configuration of the ViPLab application.

For example, the container image is most likely different for every application. The command to execute the program varies. Moreover, the parameters configured by a user (input) and the files produced by the program (output) vary. To this end, also the GUI of a ViPLab web application for running programs consists of fixed elements (such as buttons for navigation), and optional elements chosen from a predefined set of components, for example, web forms (for collecting user input and containing elements such as checkboxes, sliders, dropdown menus), a file editor, or elements for the visualization of data. 

All configurable elements of a ViPLab application are specified by a single configuration file, which we call \textit{computation template}, cf. Fig.~\ref{fig:webapp}. Computation templates are structured files in \texttt{JSON} file format\footnote{The file format is documented in the ViPLab documention~\citeW{vl0-doc-comput-template}}.
For the convenience of content creators (\eg a researcher that wants to make their software artifacts interactively explorable), ViPLab provides a web service that allows creating a computation template by means of a GUI~\citeW{vl1-create-gui}. Computation templates can also be edited manually, or programmatically in an automated process.

\begin{figure*}[htbp]
  \centering
  \includegraphics[width=0.9\textwidth]{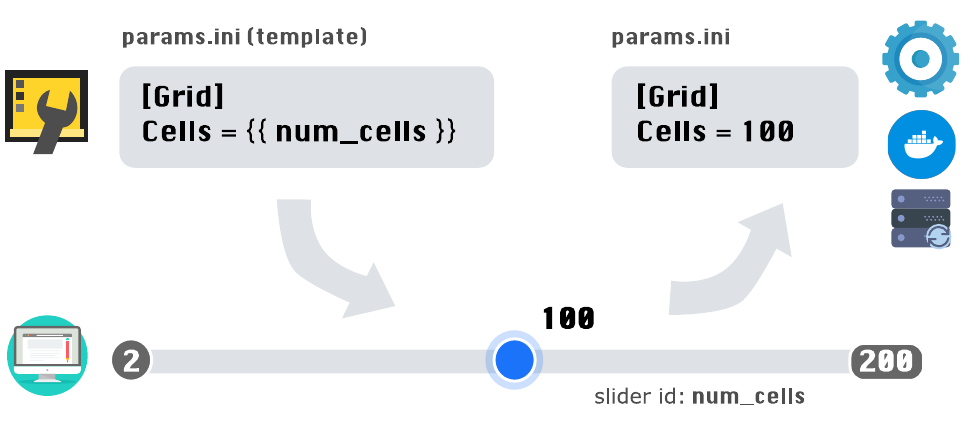}
  \caption{\textbf{Substitution mechanism to configure parameter files for research software through GUI elements in a web application.} In this example, the parameter \texttt{num\_cells} in a research application's input file is exposed in the web application via a range slider element. This is achieved by creating a \texttt{params.ini} template file which contains a placeholder token \texttt{\{\{ num\_cells \}\}} that is replaced by the current value of the range slider (here: \texttt{100}) upon submission of the computation.
  The translated input file \texttt{params.ini} is transferred to the computation backend before the computation is started.}
  \label{fig:template}
\end{figure*}

A convenient feature for the creation of web applications (following \textbf{[WA1-4]}) is a parameter substitution mechanism. An example is illustrated in Fig.~\ref{fig:template}.
The creator of a web application can provide an input file \textit{for the research software} (e.g. a parameter input file, or a mesh file) where interesting parameters have been substituted by placeholder tokens. Such a file together with one GUI element for user input per placeholder token is specified as part of the ViPLab computation template.
When a computation is triggered in the frontend, all GUI input elements are evaluated and each placeholder token is substituted back with the value of its associated element.
This constitutes a simple but effective mechanism to go from software that is configurable through input files, to software that is configurable through GUI elements in a web application.
The substitution mechanism in ViPLab can also be applied to command-line arguments to configure research applications used through a command-line interface.

\begin{figure*}[htbp]
  \centering
  \includegraphics[width=\textwidth]{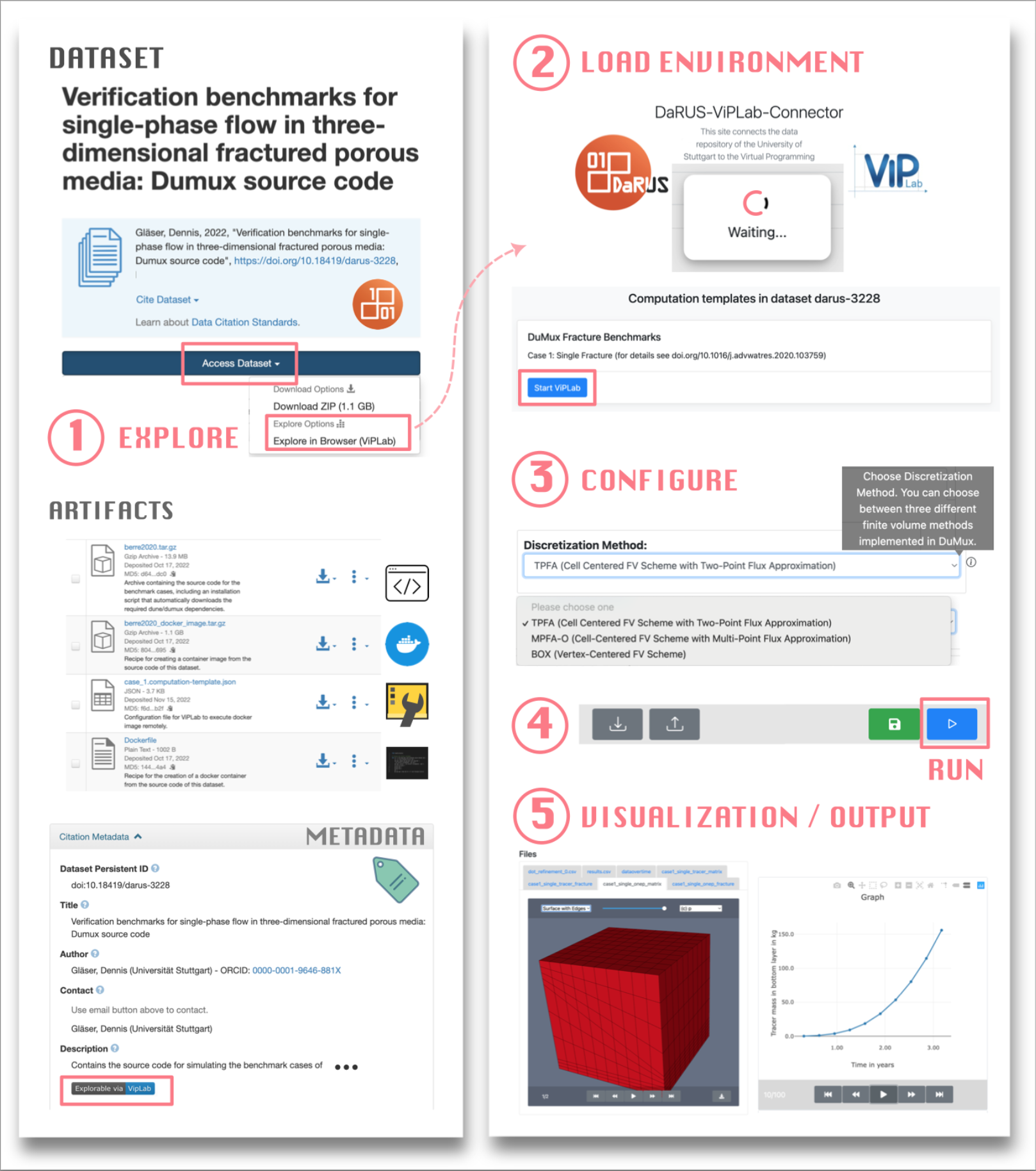}
  \caption{\textbf{Browser-based, installation-free exploration of datasets containing software artifacts via the DaRUS-ViPLab-Connector.} Left, are snapshots of a DaRUS dataset with a multi-modal representation of research software artifacts, which is cross-linked to an interactive browser-based tool for exploration in ViPLab. Right, is a workflow example for exploring an artifact via limited configurability (simple first-contact scenario). The dataset containing this example can be found at 
  \url{https://doi.org/10.18419/darus-3228}. More advanced scenarios may involve editing source code which is supported by ViPLab.}
  \label{fig:viplab-frontend}
\end{figure*}

In Fig.~\ref{fig:viplab-frontend}, we show how ViPLab can be accessed to explore an exemplary dataset in the data repository DaRUS (described in more detail in the following section). The right part in Fig.~\ref{fig:viplab-frontend} illustrates a workflow in ViPLab: (1) after starting ViPLab, (2) a computation template (describing a GUI application) is selected and the GUI application is created, (3) parameters are selected or code is edited by the user, (4) all GUI elements evaluated, the input files are created and sent to the backend, and a containerized application on the backend performs the computation and delivers results files that (5) are visualized in the web application. Currently, besides reporting output and error streams, rendering images, scientific data in \texttt{CSV} format\footnote{CSV files (comma-separated values) are assumed to contain numerical data for visualization in graphs and must be tidy, see~\cite{tidy-data}.} and \texttt{VTU}/\texttt{VTP} format\footnote{Common file formats suitable for the visualized of grid-based data used by the Visualization Toolkit (VTK)~\cite{VTK}.} can be rendered within the web application.
Additionally, all output files can be downloaded.

The computation template defining a web application for the exploration or reuse of research software artifacts is meant to be archived in the same dataset as the research software artifacts (which also includes the container image described in Section~\ref{sec:docker}). A dataset may contain more than one computation template if there are several possible applications.

% TK: I moved this down here because it's a nice conclusion
The architecture and functionality of ViPLab is described in a new online documentation~\citeW{vl}.
After the mentioned rewrite, ViPLab now meets the requirements and recommendations outlined in Section \ref{ssec:interactive-reuse} and can substantially reduce the complexity of executing research software in practice.
All components of ViPLab are published and developed as FLOSS~\citeW{viplab}.

Finally, we stress that, in fact, ViPLab was not merely adapted for suitability in the research context but generalized to support both applications: e-learning and exploring research software. The revision of all software components increased code quality and enabled the use of state-of-the-art technology. To maximize the potential user base, the designed ViPLab components are neither specific to \dumux nor to Docker.

\subsection{Multi-modal software data repository: example DaRUS}
\label{sec:darus}

The \emph{Da}ta \emph{R}epository of the \emph{U}niversity of the \emph{S}tuttgart (DaRUS) is based on the open-source research data repository software Dataverse~\citeW{b3-dataverse} and offers groups of the University (usually institutes, working groups and projects) the possibility to organize, store, share and publish their data and code.
All data published on DaRUS is discoverable via B2FIND~\citeW{b2find}, OpenAIRE~\citeW{c1-darus-openaire}, and Google Dataset Search~\citeW{google-dss}.

DaRUS uses Digital Object Identifiers (DOIs) as persistent identifiers for both datasets and files (\textbf{[MR1]}). There are no constraints on file type or size (\textbf{[MR2]}). Metadata schemes are modular and customizable and new schemes can be added (\textbf{[MR3]}). Datasets can be endowed with a license.
In addition, the multi-modal representation of research software is supported. For example, there is an API to upload files, which is especially useful for large files like container images. Finally, there is the possibility to connect external tools (plug-in system) to explore datasets and specific file types, which we use to integrate ViPLab.

The quality of published datasets is ensured through a transparent, documented publication workflow~\citeW{c0-darus-pub}.
Each dataset undergoes a quality check prior to publication, aiming for completeness and reproducibility (see \textbf{[MR5]}). However, rather than verifying these properties directly, the reviewer tries to evaluate whether the formal requirements for reproducibility are met judging from the available data.
Authors are encouraged to use open data formats that are legible in the long term and/or are common in the research domain.

In the following, we describe our adjustments and improvements toward the sustainable archival of multi-modal research software artifacts in particular.

% Metadata
In DaRUS, general information about the dataset (e.g. title, description, author, contact, subject, project, funding) and relationships to other datasets are based on the metadata standards DDI and DataCite. We enabled specific metadata regarding the use and documentation of software and code to DaRUS~\cite{darus-508_2020,darus-3291_2022} by supporting the metadata standards EngMeta~\cite{darus-500_2019} (software usage) and CodeMeta~\cite{codemeta-standard} (software development). EngMeta allows describing, for example, the research process, methods, and tools used, or the variables and parameters collected. CodeMeta allows linking to code development repositories and describing contributors, programming language, version, or software dependencies.

To support the automation of adding metadata to datasets, we developed the tool 
\texttt{Metadata2Dataverse}~\citeW{md2dv} that allows the extraction of metadata contained in (structured) metadata, parameter, or log files (software artifacts in \texttt{XML} or \texttt{JSON} format). A prominent example is the CodeMeta JSON file format. The file is ideally maintained in the code repository of the research software (for \dumux see \citeW{x1-dumux-codemeta}).
\texttt{Metadata2Dataverse} comes in form of a REST-API that converts such artifacts into a Dataverse-compatible \texttt{JSON}-based metadata file which can be uploaded to DaRUS via the regular Dataverse API.
Further working towards automation, we also implemented a GitHub Action~\citeW{b2-dataverse-action}. This can be used in an automated CI/CD workflow to convert a CodeMeta JSON file to the Dataverse-compatible format and update the metadata on a DaRUS dataset.

To enable the interactive reuse of datasets (\textbf{[MR6]}), we added the capability of linking ViPLab applications to explore datasets containing a ViPLab computation template file by building a Dataverse external tool~\citeW{externaltooldataverse}.
This so-called Dataverse-ViPLab connector (see Fig.~\ref{fig:viplab-frontend}) fulfills three tasks:
First, it translates DOIs to actual files---using user credentials when necessary---and downloads containers and computation templates.
Second, it embeds the ViPLab web application and third, it provides access to the web service for creating new computation templates.
Every dataset containing a valid ViPLab computation template (see Section~\ref{sec:viplab}), can be interactively explored directly from DaRUS (see Fig.~\ref{fig:viplab-frontend}). This is also possible before the dataset is published, allowing researchers (or RSEs) to create and test their ViPLab web applications in DaRUS.

Finally, we mention some ongoing development to enhance source code citability in DaRUS. The possibilities to cite source code are limited in DaRUS. 
Source code, as a living document, can vary substantially between
versions, but when \emph{updating} a dataset (new version), DaRUS preserves the dataset's DOI.
Moreover, source code is often published as a package (\eg\texttt{tar} archive), which means that the individual files contained in the package cannot be assigned a persistent identifier. Software Heritage~\cite{DicosmoSoftwareHeritage2017} is a repository dedicated to source code archival where persistent identifiers are available even on the code line level of each revision.
Following recommendations \textbf{[MR1]} and \textbf{[MR2]} (second part),
we added to the DaRUS publication workflow checklist~\citeW{c0-darus-pub}
that software source code is archived additionally on Software Heritage when possible and the resulting persistent identifier to the source code repository is linked on DaRUS.

\section{Pilot case involving \texorpdfstring{\dumux}{Dumux}, ViPLab, and DaRUS}
\label{sec:case_study}

As an example, we present the efforts and experiences made in the context of a recent benchmark study on single-phase flow and transport in three-dimensional fractured porous media
\citep{Berre2021Benchmarks}. We describe the steps taken in the development of a research software application for the benchmark study, which is then archived in a multi-modal representation in an institutional data repository, in a way suitable for reuse and exploration.
We conclude with an initial evaluation of the resulting dataset.

The research code example discussed in the following is part of a contribution of one research group to a benchmark study comparing the accuracy and performance of several codes from different research groups when solving specific fracture flow problems (benchmark cases), which had been defined in a preceding call for participation \citep{Berre2018participation}.
The motivation for the benchmark study is two-fold. On one hand, the report gives insights into the spread that one can expect in the solutions obtained by state-of-the-art numerical methods.
On the other hand, researchers can employ the benchmark scenarios to test newly developed methods against the solutions presented in the report, in the future. 
To this end, it is crucial that scientific peers have access to the solutions of others, and for investigations beyond the benchmark study, peers need to be able to modify and run the code. Finally, the reproducibility of the submitted benchmark results increases the trust in both the report and the software.

\paragraph{Research software}
%Regarding the research software (see Section~\ref{ssec:research-software})
The software for the contribution to the benchmark study (i.e. the research software) was implemented based on the \dumux~\citep{KochetalDumux2021} framework (see \textbf{[RS1]}). The source code was initially developed in a private Git repository by the researchers (domain experts) hosted in an institutional GitLab instance. The code repository was made publicly
accessible~\citeW{berre2020} after the publication of the study. The benchmark study compares secondary data in the form of plots through the simulation domains, and to allow for a simple (re-)production of this data, the code was developed such that all secondary data is computed automatically without requiring manual steps (see \textbf{[RS5]}, \textbf{[RS6]}). In particular, the post-processing pipeline was automated using the Python interface of the open-source visualization software ParaView~\cite{Ahrens2005ParaViewAE}. In order to facilitate the installation of the research software, the \texttt{README} file of the repository lists the basic dependencies of the software, which are usually installable via package managers in many Unix-based operating systems (see \textbf{[RS2]}). For further dependencies such as \dumux and DUNE modules, an installation script is provided to facilitate the setup of a suitable production environment (see \textbf{[RS3]}).

\paragraph{Containerized application}
The repository furthermore contains a \texttt{Dockerfile} that describes the realization of a productive environment reusing the installation script (see \textbf{[RS7]}, \textbf{[CP1]}). Note that both the installation script and the \texttt{Dockerfile} were created automatically with software tools provided by \dumux, which had been implemented by an RSE in the scope of a software infrastructure research project, cf. Section~\ref{ssec:dumux}.
The researchers uploaded a pre-built container image to the container registry of the GitLab repository such that the installation procedure may be skipped in a reuse scenario (see Section~\ref{sec:docker}).

\paragraph{Software data repository}
All research software artifacts have been published on the institutional repository DaRUS (see~\ref{sec:darus}). To this end, a small modification to the developed \texttt{Dockerfile} was necessary. Up to this point, the \texttt{Dockerfile} cloned the code from the Git repository and then executed the installation script also provided in that repository accessed via a possibly non-persistent URL.
This was changed such that the source code published in the DaRUS dataset~\citeW{b1-dataset-code} is used instead (see \textbf{[RS4]}). This source code version corresponds to the version used in the prebuilt container image and is accessible via a persistent identifier (DOI). Moreover, the institutional repository DaRUS fulfills our requirements on long-term availability. Alongside the source code, the dataset also contains the \texttt{Dockerfile}, and a compressed \texttt{tar} archive of the pre-built container image (see \textbf{[MR2]}). The metadata required by the curated repository and enforced by the infrastructure provider in a review process (see \textbf{[MR5]}) was added manually. 

\paragraph{Browser-based access}
Four different ViPLab computation template files (see \textbf{[WA6-7]}) were created and added to the dataset. Each computation template corresponds to a separate benchmark scenario. The applications were configured such that only a few simple parameters can be changed (see \textbf{[WA1]}, \textbf{[WA3]}) such as the resolution of the computational grid and the spatial discretization method in case of the first scenario, for instance. The computation templates were developed in collaboration with the infrastructure provider and a research software developer, and they required a small modification to the container image: a script that takes a benchmark case number as input and calls the corresponding pre-built executable application available in the container was developed and set as the \emph{entry point} of the container.

\paragraph{Result data repository}
In addition to the software dataset, the primary simulation results and the secondary data of the first benchmark case, as submitted to the study by the research group and
referenced in the journal publication\cite{Berre2021Benchmarks}, have been published as a separate DaRUS dataset~\citeW{b0-dataset-data}.
Since the default parameters in the computation templates mentioned above were chosen to match the parameters used in the original benchmark study (see \textbf{[WA2]}), one can reproduce these results conveniently in the browser.

\paragraph{Possible reuse scenarios}
As the data is provided in various forms, a variety of reuse scenarios is possible.
Interested users may start by accessing the browser-based application that
can be launched from the code publication (see Fig.~\ref{fig:viplab-frontend}).
This allows the reproduction of some (but not all) of the data presented in \citet{Berre2021Benchmarks}. Additionally, users may choose parameters
that go beyond the investigations of the original study. However, the choice is limited to those parameters exposed by the creator of the application.

For investigations beyond this initial access with limited configurability, users can download and launch the provided container image containing the prebuilt applications. This exposes many more runtime parameters but also requires more knowledge about the program. The container can also be used to change source code from within a working environment, hence, providing flexibility without additional effort.

However, a user may want to alter the build environment to see if this affects the results or to adapt the software artifact to their workflow environment. As the \texttt{Dockerfile}, installation scripts, and a \texttt{README} file are provided in the dataset, it is documented how a suitable environment may be instantiated. This data can also become necessary if, for instance, the provided pre-built container image can no longer be executed on a newer platform. Let us give an example: initially, the C++ code in the project had been compiled with optimization options targeted for a specific CPU architecture. (This is common to minimize the execution time of computation-heavy applications.) When trying to run the Docker image on a different computer and by another group member, the program failed to execute correctly. Using the \texttt{Dockerfile} and installation scripts, the installation could easily be modified to exclude the optimization flag. The container image was rebuilt to include the portability improvement. Between creation and testing, time passes. The example illustrated the importance of supplying additional data (\texttt{Dockerfile} and installation scripts).

Finally, we summarize how this pilot case illustrates the interaction between various roles defined in Section~\ref{ssec:roles}.
The \dumux source code for the simulation of the benchmark cases was developed by \textit{researchers}.
For the creation of the installation script and \texttt{Dockerfile}, they reused the automation scripts developed by a \textit{research software engineer}.
The \textit{infrastructure providers} (who initially implemented a suitable metadata scheme) assisted the \textit{researchers} in annotating their dataset.
Moreover, \textit{infrastructure providers} provide computing resources and developed and maintain the hosting service. The web application infrastructure was developed by a \textit{research software engineer} in close collaboration with \textit{infrastructure providers}. The data repository (and the cross-linked web application) serves as a point of first contact for both \textit{scientific peers} and \textit{non-academic users} to explore the software.

\section{Discussion and open challenges}

The approach presented in this work yields \textit{FAIR} research software artifacts that are reusable in a large variety of ways, while little additional work is required to comply with it, under the following prerequisites:
\begin{enumerate} 
    \item the pipeline for the computation of results is fully automated, 
    \item the software is built on a framework that provides utilities for generating installation scripts and/or container images,
    \item the software relies on a portable build system,
    \item the software has no dependency on proprietary or closed-source products,
    \item open licenses for all research software artifacts.
\end{enumerate}

Consequently, these items can also be considered best-practice recommendations for sustainable archival and reuse.
In our experience with the described pilot implementation at the University of Stuttgart and an associated hands-on project workshop with 19 participants, we found that these prerequisites are often not given in practice. (The participants arrived with their own so-far developed workflow.) If this is the case, a substantial amount of work---at least initially---may be necessary to follow the described approach.
Particularly, we encountered significant obstacles with software distributed without a portable build system or requiring proprietary software products at any stage. Hence, in order to create a \textit{FAIR} research software culture, we recommend to all research software developers to make it a main priority to create free, open-source workflows with trustworthy, portable software (FLOSS) in stable development environments. The use of graphical user interfaces in the workflow led to manual instead of automated steps. Therefore, we recommend tools with scripting capabilities for automation and command-line interfaces.

In summary, a major challenge is the \textit{automation} of the archival process. It has to begin with the workflow itself, and therefore, right from the beginning of the research activity. While some work overhead can be significantly reduced for individual researchers following some of the mentioned best practices (e.g. the automated creation of installation recipes within a research software framework, or the automated creation of containerized applications suitable for integration with a browser-based application,) open questions remain. For example, to better describe the research software artifact, the software itself has to be adapted to export metadata that is otherwise not accessible. This metadata has to be compatible with what is importable to the data repository. Moreover, the typical research software installation process (e.g. \texttt{git clone}, or \texttt{pip install}) might differ from the installation process relying on software accessed from data archives designed for long-term storage (e.g. Software Heritage's API), which requires adaption of already established workflows.

Due to special hardware and software requirements for high-performance computing and associated long run times and costs for reproduction, such applications require additional concepts, see e.g.~\cite{Wagner2022}.
But even for smaller applications, the \textit{cost} associated with providing easy-to-use browser-based \textit{reuse and exploration} can become an issue. Firstly, it has to be evaluated when interactive reuse makes sense and when it does not: do I want to wait hours for results? Secondly, as a researcher investing time (paid by public funding) to make software artifacts reusable, it is natural to expect that other parties will be able to reuse the artifacts freely. Scalable solutions that benefit everyone in society but adhere to the FAIR principles would have to be decentralized~\cite{Hanke2021}. That means each user of the system provides computing resources themselves (and although commercial resources such as cloud infrastructure subscriptions can be cost-effective, they must remain optional). Naturally, software provenance plays an integral role in developing secure solutions.

How to best describe software artifacts FAIRly in practice is a challenge.
Which kind of \textit{documentation and metadata} is essential and will be essential for future reuse? Which modes of data are most important to assure findability, accessibility, interoperability, and reuse? What is the ideal form of representing an interactive, extensible software artifact by metadata given that future use cases might be difficult to predict by its creator? 

We believe that workable solutions can be achieved in a collaboration of infrastructure providers, scientists, and research software engineers and that
ultimately, reusable and extensible software artifacts have tremendous potential for increasing the productivity, efficiency, and sustainability of scientific advances.

\begin{acknowledgements}
Funded by Deutsche Forschungsgemeinschaft (DFG, German Research Foundation) - project numbers 391049448; 442146713; and, under Germany's Excellence Strategy, - EXC 2075 – 390740016.
T. Koch acknowledges funding from the European Union’s Horizon 2020 Research and Innovation program under the Marie Skłodowska-Curie Actions Grant agreement No 801133.
D. Gläser and A. Seeland would like to thank the Federal Government and the Heads of Government of the Länder, as well as the Joint Science Conference (GWK), for their funding and support within the framework of the NFDI4Ing consortium.
B. Flemisch and S. Hermann acknowledge the support of the Stuttgart Center for Simulation Science (SimTech).
We would like to thank James Hartill for developing a web application prototype for \dumux~\citeW{dwebapp} that allowed us to practically test and explore such technology early on. Moreover, we thank Dorothea Iglezakis for her constructive comments on the initial manuscript.
\end{acknowledgements}

\section*{CRediT contributor statement}
{\small
\textbf{Timo Koch:}  Conceptualization; Writing – Original Draft Preparation; Software; Visualization (lead); Data Curation.
\textbf{Dennis Gläser:} Writing – Original Draft Preparation; Software; Visualization (supporting); Data Curation.
\textbf{Anett Seeland:} Writing – Original Draft Preparation; Data Curation; Software.
\textbf{Katharina Schulze:} Writing – Original Draft Preparation; Software.
\textbf{Sarbani Roy:} Writing – Original Draft Preparation; Software.
\textbf{Kilian Weishaupt:} Writing – Review \& Editing; Software.
\textbf{David Boehringer:} Writing – Review \& Editing; Supervision; Resources.
\textbf{Sibylle Hermann:} Conceptualization; Supervision; Writing – Original Draft Preparation.
\textbf{Bernd Flemisch:} Conceptualization; Supervision; Writing – Original Draft Preparation; Software.
}

\section*{Conflict of interest statement}
{\small
T.K., D.G., S.R., K.W., and B.F. are developers and maintainers of the \dumux project, and T.K. is a core developer of DUNE. K.S., A.S., and D.B. are developers and maintainers of the ViPLab project, A.S., and S.H. are developers and maintainers of the DaRUS infrastructure service.
}

\section*{Abbreviations}
{\small
\textbf{API:} application programming interface;
\textbf{CI/CD:} continuous integration and continuous delivery; \textbf{DOI:} Digital Object Identifier; \textbf{FAIR:} findable, accessible, interoperable, reusable, following the FAIR principles for data~\cite{Wilkinson2016FAIR} or software~\cite{Barker2022}; 
\textbf{FLOSS:} free/libre open source software; \textbf{GUI:} graphical user interface; \textbf{OCI:} Open Container Initiative~\citeW{oci}; \textbf{RSE:} research software engineer; \textbf{URL:} uniform resource locator.
}

\bibliographystyle{unsrtnat}      % basic style, author-year citations
\bibliography{main}   % name your BibTeX data base

{\footnotesize%
\bibliographystyleW{unsrtnat}
\bibliographyW{web}
}

%\begin{thebibliography}{}
% etc
%\end{thebibliography}

\end{document}